\documentclass[english,prb,twocolumn,amsmath,amssymb,floatfix,bibnotes,footinbib,citeautoscript,superscriptaddress]{revtex4-1} 
\usepackage[T1]{fontenc} 
\usepackage[latin9]{inputenc} 
\usepackage{geometry}   \synctex=-1 
\usepackage{amstext} 
\usepackage{graphicx} 
\usepackage{esint}
\usepackage{amssymb,amsmath}
\usepackage{placeins}
\usepackage{color}
\usepackage{feynmf}
\usepackage{epstopdf}

\makeatletter

 \@ifundefined{textcolor}{} {

\definecolor{BLACK}{gray}{0} 
\definecolor{WHITE}{gray}{1} 
\definecolor{RED}{rgb}{1,0,0} 
\definecolor{GREEN}{rgb}{0,1,0} 
\definecolor{BLUE}{rgb}{0,0,1} 
\definecolor{CYAN}{cmyk}{1,0,0,0} 
\definecolor{MAGENTA}{cmyk}{0,1,0,0} 
\definecolor{YELLOW}{cmyk}{0,0,1,0} }

\pdfoutput=1

\usepackage[colorlinks=true,citecolor=blue,linkcolor=magenta]{hyperref} 
\usepackage{amsmath} 
\usepackage{graphicx} 
\usepackage{dsfont} 
\usepackage{changepage} 
\usepackage{appendix}

\usepackage[T1]{fontenc}

   \global\long 
\def\sp{,\qquad}     \global\long 
\def\ex#1{\left\langle #1\right\rangle }     
\def \bea{\begin{eqnarray}}
\def \eea{\end{eqnarray}}
\def \beq{\begin{equation}}
\def \eeq{\end{equation}}

                     \global\long 
\def\II{\mathcal{I}}     \global\long 
\def\OO{\mathcal{O}}

\def\d{\mathrm{d}} 
\def\pdr{
\partial}

\@ifundefined{showcaptionsetup}{}{

\PassOptionsToPackage{caption=false}{subfig}} 
\usepackage[caption=false]{subfig}
\makeatother

\usepackage{babel}

\begin{document}

\title{Fate of the One Dimensional Ising Quantum Critical Point Coupled to a Gapless Boson}

\author{Ori Alberton}
\affiliation{Department of Condensed Matter Physics, Weizmann Institute of Science, Rehovot 76100, Israel}
\affiliation{Institute of Theoretical Physics, University of Cologne, D-50937 Cologne, Germany}
\author{Jonathan Ruhman}
\affiliation{Department of Physics, Massachusetts Institute of Technology, Cambridge, MA 02139, USA}
\author{Erez Berg}
\affiliation{Department of Condensed Matter Physics, Weizmann Institute of Science, Rehovot 76100, Israel}
\author{Ehud Altman}
\affiliation{Department of Condensed Matter Physics, Weizmann Institute of Science, Rehovot 76100, Israel}
\affiliation{Department of Physics, University of California, Berkeley, CA 94720, USA}


\begin{abstract}
The problem of a quantum Ising degree of freedom coupled to a gapless bosonic mode appears naturally in many one dimensional systems, yet surprisingly little is known how such a coupling affects the Ising quantum critical point. We investigate the fate of the critical point in a regime, where the weak coupling renormalization group (RG) indicates a flow toward strong coupling. Using a renormalization group analysis and numerical density matrix renormalization group (DMRG) calculations we show that, depending on the ratio of velocities of the gapless bosonic mode and the Ising critical fluctuations, the transition may remain continuous or become fluctuation-driven first order. The two regimes are separated by a tri-critical point of a novel type.
\end{abstract}

\maketitle

\section{Introduction}

The problem of a quantum Ising transition occurring in a system with gapless phonon modes naturally emerges as the low energy effective theory in a variety of physical one dimensional systems. Examples include structural transitions of dipolar particles in an elongated trap \cite{Ruhman2012} and in ion traps \cite{Fishman2008,Shimshoni2011,Podolsky2014,Giorgi2010}, one dimensional Fermi gas with attractive interactions \cite{Ruhman2015}, coupled bosonic chains \cite{Orignac1998,Lecheminant2012}, insulator-metal transition induced by a tilted magnetic field in graphene \cite{Murthy2014}, and spin polarized electrons in a one dimensional quantum wire near a conductance plateau transition \cite{Meyer2007,Sitte2009,Mehta2013,Sun2016}. This transition also occurs in certain lattice models of fermions \cite{Bauer2013}. At the transition, the symmetry of the system is dynamically enlarged to a full Lorentz symmetry \cite{Sitte2009} or, under certain conditions, even to an emergent supersymmetry \cite{Huijse2015}.

The coupling between the gapless phonon and the Ising degree of freedom poses a challenge in understanding the low energy behavior of the system, especially near the critical point where the Ising excitations become gapless as well. Integration over the gapless Ising degrees of freedom in order to obtain an effective action for the bosons is unfeasible as it leads to logarithmic divergencies. On the other hand, one cannot simplify the problem by integrating out the bosons, as this will generate non-local interactions for the Ising degrees of freedom that are then difficult to deal with.

This problem was treated using weak coupling one-loop renormalization group (RG) analysis in a previous work \cite{Sitte2009}. It was found that there is a regime in which the coupling to the phonon is marginally irrelevant in the RG sense, therefore the two sectors effectively decouple at large length scales. In this case the transition is still described by an Ising like critical point albeit with strong logarithmic corrections. However, there is another regime in which the weak coupling RG leads to a runaway flow away from the Ising critical point towards strong coupling. The fate of the transition in this latter regime is not known. 

\begin{figure}
	\includegraphics[scale=0.5]{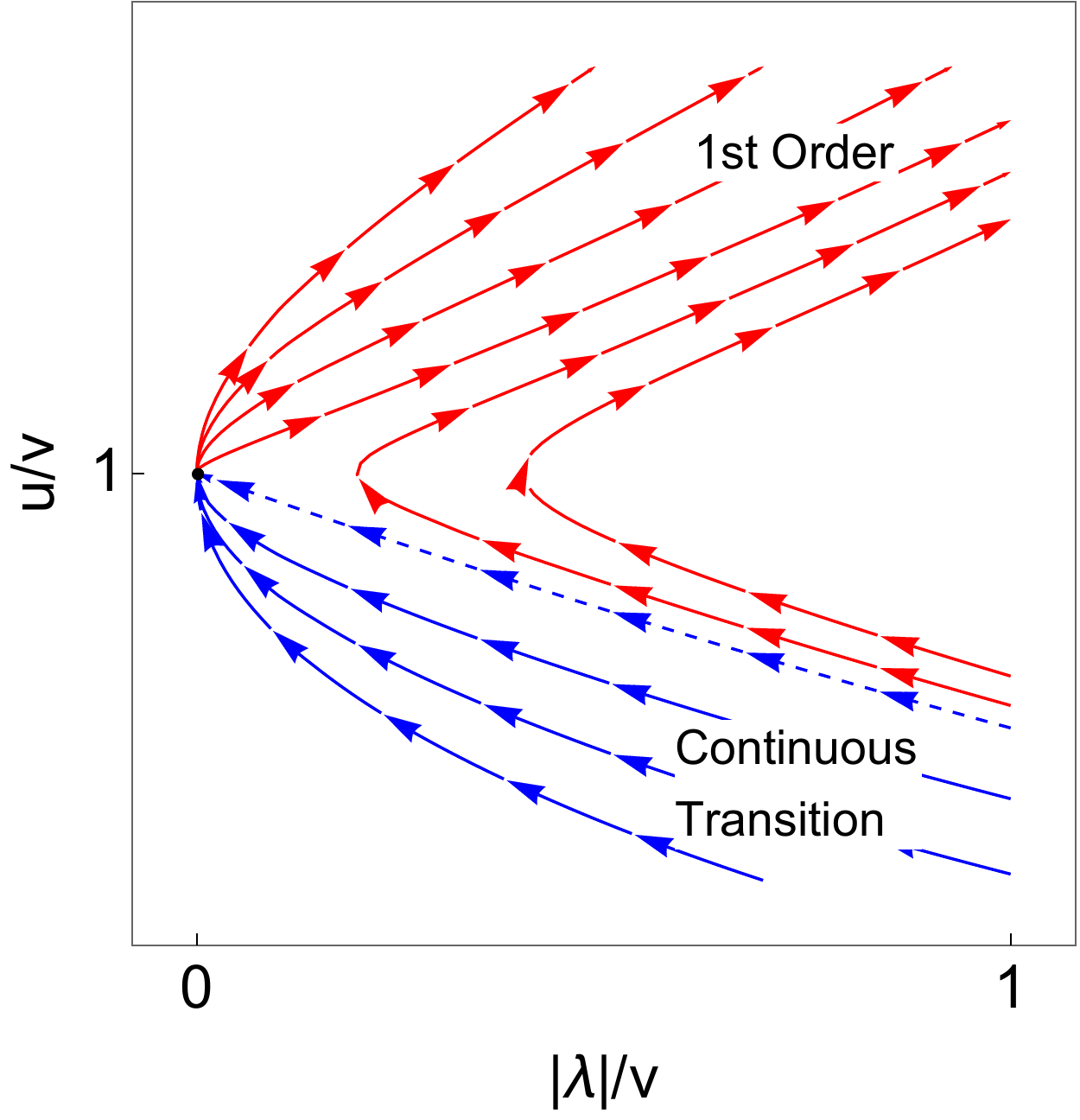} 
	\caption{\label{fig:Flow-diagram}(Color online) Schematic RG flow diagram on the critical manifold $\Delta=0$ close to the $u/v=1,\lambda=0$ fixed point, determined from Eqs. (\ref{eq:du/dl-1})-(\ref{eq:dg/dl-1}) , where $f\left(u/v=1\right)>0$. The dashed blue line is the separatrix which separates the weak coupling regime and the strong coupling regime. In the following scetions we show that the strong coupling regime corresponds to a first order transition.} 
\end{figure}

In this paper we study the transition in the regime where the weak coupling RG flow is towards strong coupling. Using RG calculations and numerical density matrix renormalization group calculations (DMRG), we establish the critical behvior shown in Fig. \ref{fig:Flow-diagram}. The result of the RG flow to strong coupling is a region  
 showing phase separation. There, quantum fluctuations destabilize the critical point leading to separation into regions of different densities where the Ising degree of freedom is gapped (Fig. \ref{fig:schamatic}).    
 
The rest of this paper is organized as follows. In section \ref{sec:RG analysis} we introduce the model and the resulting RG flow diagram. In section \ref{sec:RG-Flow} we perform a mean field calculation  which suggests that the Ising critical point can become unstable towards phase separation. We then use the one-loop RG equations to show that the compressibility diverges at a finite scale in the strong coupling regime. In section \ref{sec:dmrg} we investigate this model numerically using DMRG and confirm the emergence of phase separation. Finally section \ref{sec:summary} provides summary and discussion of the results.

\section{RG analysis\label{sec:RG analysis}}

The theory describing gapless phonon excitations coupled to Ising degrees of freedom near criticality is captured by the following low-energy effective model: 
\begin{eqnarray}
	& H_{\text{L.L}}=\frac{v}{2\pi}\int\d x\left[K\left(\pdr_{x}\theta\right)^{2}+\frac{1}{K}\left(\pdr_{x}\phi\right)^{2}\right]\label{eq:H_LL}\\
	& H_{\text{Ising}}=\int\mathrm{d}x\left[\frac{iu}{2}\left(\xi_{L}\pdr_{x}\xi_{L}-\xi_{R}\pdr_{x}\xi_{R}\right)-i\Delta\xi_{L}\xi_{R}\right]\label{eq:Ising H}\\
	& H_{\text{Int}}=\int\d x\left[-\frac{\lambda}{\pi}\left(\pdr_{x}\phi\right)i\xi_{L}\xi_{R}\right]\label{eq:Coupling H}. 
\end{eqnarray}
The Luttinger liquid Hamiltonian, Eq. (\ref{eq:H_LL}), describes gapless phonon degrees of freedom with $K$ the Luttinger parameter and $v$ the phonon velocity. The density modulation with respect to the average density is $-\frac{1}{\pi}\pdr_{x}\phi=\delta\rho(x)$, and $\theta$ is the phase conjugate to it ($\frac{1}{\pi} vK\pdr_{x}\theta$ is the current density). 

The low-energy excitations of the Ising degrees of freedom near the critical point are captured by the single non-chiral Majorana mode of the Hamiltonian in Eq. (\ref{eq:Ising H}) with right- and left-moving Majorane field, $\xi_{R,L}$ \cite{Gogolin2004}. The Majorana mass $\Delta$ tunes the distance from the critical point and $u$ is the velocity at the critical point. When $\Delta>0$ the ground state of $H_{\mathrm{Ising}}$ exhibits long range Ising order and breaks the $\mathbb{Z}_{2}$ symmetry, while $\Delta<0$ corresponds to the quantum disordered phase.

Finally, $H_{\mathrm{int}}$ in Eq. (\ref{eq:Coupling H}) is the simplest coupling term between the two sectors. Because this coupling  respects both the $\mathbb{Z}_2$ symmetry of the Ising sector and the $U(1)$ symmetry of the phonons, it should appear in a generic case where an Ising transition occurs in a system with a scalar bosonic mode. Note, however, that both the Ising theory and the Luttinger liquid model are separately Lorentz invariant, and the coupling term breaks the Lorentz symmetry.

The effect of the coupling term on the critical properties of the model above has been previously studied using a perturbative one loop RG analysis \cite{Sitte2009}, controlled by the parameter 
\begin{equation}
	g \equiv \frac{K\lambda^{2}}{uv}.
\end{equation}
The RG step consists of perturbatively integrating out modes in a momentum shell $(\Lambda/b,\Lambda)$, where $\Lambda$ is the UV momentum cutoff of the theory. Momentum is rescaled by a factor $b$ and frequency by the factor $b^{z}$, with $z$ the dynamical exponent. The fields $\phi,\xi_{R/L}$ (in momentum/frequency space) are rescaled by factors $\sqrt{Z_{B}}\equiv b^{\eta_{B}/2},\sqrt{Z_{F}}\equiv b^{\eta_{F}/2}$ respectively, chosen such that the parameters $v,K$ and the unit prefactor of the dynamical term in the fermion action are invariants of the flow. Note that this implies rescaling of the space-time fields with $\sqrt{\tilde{Z}_{B/F}}=b^{-1-z} \sqrt{Z_{B/F}}$.

The one loop RG flow on the critical manifold $\Delta=0$ has two distinct regimes. When $u/v<1$, the coupling term is marginally irrelevant and the flow is towards a fixed point at $u/v=1$, $\lambda=0$. In this case the fixed point is described by the Lorentz invariant theory of three massless Majorana fermions with equal velocities. However, when $u/v>1$ the flow is towards strong coupling where the weak coupling RG eventually breaks down. The fate of the transition in this strong coupling regime is the main focus of this paper.

A peculiar feature of the one loop flow derived in Ref. [\onlinecite{Sitte2009}] is the existence of a line of fixed points at $u/v=1$ for all $\lambda$, separating the strong coupling regime and the regime in which the coupling is irrelevant. To determine if this fixed line is generic and not just an artifact of the one loop calculation we now consider corrections to the RG flow from two loop diagrams. The one loop scaling equations derived in Ref. [\onlinecite{Sitte2009}] supplemented by the leading two-loop corrections must be of the general form:
\begin{align}
	\begin{split}
	\frac{\partial \left( u/v \right)}{\partial\ell} =  -\frac{K}{\pi^{2}}\left (\frac{\lambda}{v} \right)^2 &\left[\frac{1}{\left(1+u/v\right)^{2}}-\frac{1}{4u/v}\right]\frac{u}{v} \\
	&+f\left(u/v\right)\left(\frac{\lambda}{v} \right)^4
	\end{split} \label{eq:du/dl-1} \\
	\frac{\partial\left( \lambda/v \right)}{\pdr\ell} = -\frac{K}{2\pi^{2}}\left(\frac{\lambda}{v}\right)^3 &\left[\frac{1}{u/v\left(1+u/v\right)}\right. \nonumber\\
	&\left.\quad+\frac{1}{\left(1+u/v\right)^{2}}-\frac{3}{4u/v}\right]\label{eq:dg/dl-1},
\end{align}
where $\ell = \log(b)$, and $f$ some function of the velocity ratio. The $\OO\left(\lambda^{4}\right)$ correction originates from two loop diagrams contributing to the fermion and phonon self energies. 

We do not expect $f\left(u/v=1\right)$ to vanish since the Hamiltonian does not have any special additional symmetries when $u=v$ and $\lambda\ne0$. Therefore, in order to determine how this term affects the putative fixed line $u/v=1$ it is enough to know the sign of $f\left(u/v=1\right)$. This can be done by explicitly calculating the two-loop diagrams that contribute to the $\beta$ functions \eqref{eq:du/dl-1},\eqref{eq:dg/dl-1}. Here, however, we will only provide an argument for the sign of $f(u/v)$.
If $f\left(u/v=1\right)<0$ this correction creates flow lines which start flowing away from the fixed point $(u/v=1, \lambda=0)$ but eventually return to it in a cyclic manner. This type of flow is impossible, as it violates the C-theorem according to which the central charge should decrease monotonically under the RG flow \cite{Zamolodchikov1986}. We conclude that the two-loop correction is likely to be positive, $f\left(u/v=1\right)>0$, leading to a flow diagram of the general form shown in Fig. \ref{fig:Flow-diagram}. Hence with addition of the two loop correction the line of fixed points is likely to be eliminated. In its place there is a more conventional (tri) critical flow line that separates between the flow to weak and strong coupling.

\section{Divergent Compressibility\label{sec:RG-Flow}}

Before investigating the RG flow in more detail, we gain intuition into the behavior of the system in the strong coupling limit by treating the coupling term within a mean field approximation. To this end we decouple the interaction term as follows

\begin{eqnarray}
	H_{\text{MF}} & = & H_{\text{LL}}+H_{\text{Ising}}\left(\Delta=0\right)\nonumber \\
	& & -{\lambda\over \pi}\int\d x\left(\Phi\pdr_{x}\phi+\Delta_{\phi}i\xi_{L}\xi_{R}\right)\label{eq:H_mf}.
\end{eqnarray}
Here the mean field parameters $\Phi,\Delta_{\phi}$ obey the self consistency conditions 
\begin{equation}
	\Phi=\ex{i\xi_{L}\xi_{R}}_{MF}\sp\Delta_{\phi}=\ex{\pdr_{x}\phi}_{MF}\label{eq:mf conditions}. 
\end{equation}

\begin{figure}
\includegraphics{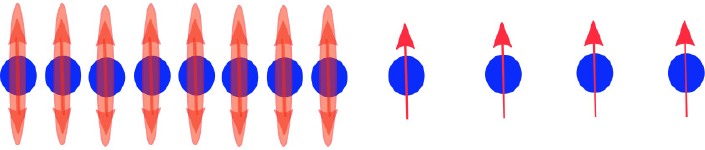} 
\caption{\label{fig:schamatic}(Color online) Schematic illustration of the system separated into a region with higher density and no Ising order, and a lower density region with Ising order. This will be the situation when the coupling $\lambda$ is positive, if it is negative the Ising degree of freedom will order (disorder) in higher (lower) density regions.}
\end{figure}

It is  straight forward to obtain the self consistent solution for the density shift
\begin{equation}
	\ex{
	\partial_{x}\phi}_{\text{MF}}=\pm\frac{\pi u}{\lambda}\Lambda\left[\sinh\left(\frac{2\pi^{2}uv}{K\lambda^{2}}\right)\right]^{-1}\label{eq:mf result}. 
\end{equation}

The non-zero mean field value of this shift implies a divergent compressibility and instability toward phase separation. If the overall density is fixed, the system separates into regions having the negative density shift and others with the positive shift. The density shift also gives a mass to the Ising sector through the linear coupling term, thus taking the Ising sector away from criticality. As illustrated schematically in Fig. \ref{fig:schamatic}, the phase separated regions will be Ising ordered and disordered, respectively, rather than critical.

The mean field analysis cannot be entirely correct, however, as it predicts that the critical point is always unstable regardless of the ratio $u/v$. From the RG analysis of Ref. [\onlinecite{Sitte2009}] we know that there is a regime ($u/v<1$), where the coupling $\lambda$ is irrelevant and the Ising criticality is recovered asymptotically at long wavelengths.
In this regime competing divergent fluctuations in other channels invalidate the mean field theory. 
Because the mean field theory does not treat all the fluctuations on the same footing it is important to verify its result also in the regime $u/v>1$. 

We will look for signatures of the predicted phase separation using the weak coupling RG. A clear indication of phase separation would be divergence of the  compressibility.  
The latter can be obtained from the $k\to0$ limit of the static density-density correlation function 
\bea
\chi\left(k,g\right)&\equiv&\frac{1}{\pi^{2}}\int \mathrm{d}x \mathrm{d}\tau e^{ikx}C(x,\tau,g) \nonumber\\
C(x,\tau,g)&\equiv&\ex{\partial_{x}\phi(x,\tau)\partial_{x}\phi(0)},
\eea
In this notation we emphasize that $\chi$ should depend on the value of the dimensionless parameter $g$. The real space correlation function must obey the scaling relation $C(x,\tau,g)=\tilde{Z}_B b^{-2} C(x/b,\tau/b^z,g(b))$
to ensure invariance of the physical correlations under the RG transformation. Upon performing the Fourier transform 
we can write the scaling relation obeyed by the static density correlation function in momentum space
\bea
	\chi\left(k,g_{0}\right) &=&  e^{-3\ell}e^{\int_0^\ell{\left[-z(\ell')+\eta_{B}(\ell')\right]\mathrm{d}\ell'}}\chi\left(e^\ell k,g\left(\ell\right)\right)  \nonumber\\
	&=&\exp\left[{K\over 2\pi^2 v}\int_0^\ell {\lambda^2\over u} d\ell'\right]\chi(e^\ell k,g(\ell)),
	\label{eq:compressibility scaling}
\eea
where we used the values 
\begin{align}
	z(\ell)  & =  1+\frac{K\lambda(\ell)^2}{4\pi^2u(\ell)v} \label{eq: z}\\
	\eta_{B}(\ell) & =  4+\frac{3K\lambda(\ell)^{2}}{4\pi^{2}u(\ell)v} \label{eq: eta_B}
\end{align}
calculated from the RG flow \cite{Sitte2009} (for details see Appendix \ref{sec:Calculation of eta_B and z}).
Here $g_{0}\equiv g\left(0\right)$ is the value of $g$ in the microscopic scale and $\ell=\log\left(b\right)$.  

Eq.(\ref{eq:compressibility scaling}) can be used to compute the compressibility of a system of a large finite size $L$, $\kappa_L(g_0)=\chi(2\pi/L,g_0)$. The scaling relation connects this, still unknown, compressibility to the easily calculable compressibility of a renormalized system of minimal size. In other words, by running the flow to the point that the running cutoff $\Lambda_0 e^{-\ell}$ reaches the infrared cutoff $\Lambda_{IR}=2\pi/L$ we end up with a renormalized Hamiltonian defined on a single microscopic grain of size $a$:
\begin{equation}
	H_{\text{eff}}=\frac{1}{2}\kappa_{0}^{-1}\left(\delta\rho\right)^{2}-\lambda_{\text{eff}}\delta\rho\left(i\xi_{L}\xi_{R}\right)+\Delta_{FS}\left(i\xi_{L}\xi_{R}\right), 
\end{equation}
where $\kappa_{0}=K/\pi v$, $\lambda_{\text{eff}}$ is a renormalized effective coupling and $\Delta_{FS}$ is the finite size gap.  The compressibility of this renormalized system can be read off directly from the first term of $H_{\text{eff}}$, namely $\kappa_a(g_{IR})=\kappa_0$. To connect the compressibility $\kappa_L(g_0)$ to $\kappa_0$ we need to compute the integral in (\ref{eq:compressibility scaling}) with an upper limit $\ell_L=\ln(L/a)$. This is done by a change of variables using Eq. (\ref{eq:du/dl-1}) but neglecting the two loop correction
\beq
\int_0^\ell {\lambda^2\over u} d\ell'={\pi^2\over K}\int_{u_0}^{u(\ell)}{1\over u^2}\left({1\over 4uv}-{1\over(u+v)^2}\right)^{-1}.
\eeq
The last integral can be evaluated to give
\begin{eqnarray}
	\kappa\left(L,g_{0}\right) & = & \left(\frac{u\left(\ell_{L}\right)}{u_{0}}\right)^{1/2}\nonumber \\
	& & \times\exp\left[-\frac{2v}{u\left(\ell_{L}\right)-v}+\frac{2v}{u_{0}-v}\right]\kappa_{0}\label{eq:comp as function of u}, 
\end{eqnarray}
where $u_{0}$ is the Ising velocity in the microscopic scale.

The flow $u(\ell)$ is calculated in Appendix \ref{sec:Calculation-of l*}. When $u_{0}/v>1$ the renormalized velocity diverges at a finite scale $\ell^{*}$ given by 
\begin{equation}
	\ell^{*}=\frac{2\pi^{2}v}{15K\lambda_{0}^{2}u_{0}}\frac{\left(15u_{0}^{3}+5u_{0}^{2}v+5u_{0}v^{2}-v^{3}\right)}{\left(u_{0}-v\right)}.
\end{equation}
This implies a divergent compressibility for a system size larger than $L^{*}\sim ae^{\ell^{*}}$, indicating an instability towards phase separation at the scale $L^{*}$. 

When $u_{0}\gg v$ we can write an explicit expression for $u\left(\ell\right)$. In this regime the compressibility diverges as 
\begin{eqnarray}
	\chi & \sim & \left(1-\frac{\II}{2\pi^{2}}\frac{u_{0}^{2}}{v^{2}}\log\left(\frac{L}{a}\right)\right)^{-1/4}, 
\end{eqnarray}
where $\II\equiv\lambda_{0}^{2}Ku_{0}v/(u_{0}-v)^{4}$ is a constant under the RG flow.

When $u_{0}/v<1$ the flow is towards the fixed point $u=v$, that is $u\left(\ell\to\infty\right)=v$. Note that this means that the compressibility diverges in the limit $\ell_{IR}\to\infty$ due to the exponent in Eq. (\ref{eq:comp as function of u}). However, the divergence in this case is slow 
\begin{equation}
	\chi\sim\exp\left(2\left(\frac{5\mathcal{I}}{16\pi^{2}}\log\left(\frac{L}{a}\right)\right)^{1/5}-\frac{2v}{v-u_{0}}\right)\label{eq:chi close to fixed point},
\end{equation}
so the fact that the compressibility diverges at the fixed point itself will not affect any finite size system in the $u/v<1$ case.

\section{dmrg analysis\label{sec:dmrg}}
\begin{figure*}
	\begin{centering}
		\subfloat[\label{fig:dens+spin J=1.7 not critical}$J=1.7$]{\includegraphics[scale=0.36]{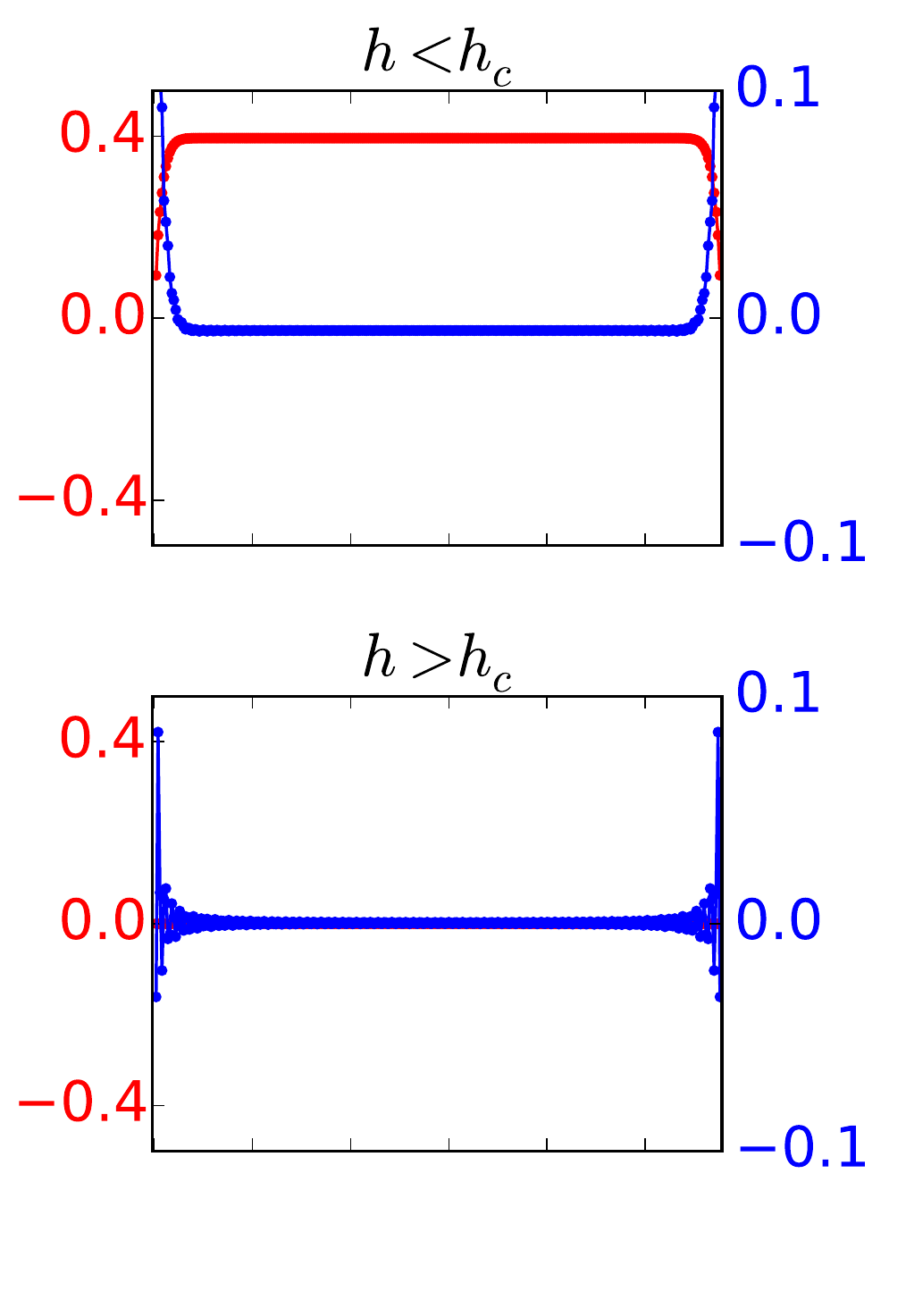} }
		\subfloat[\label{fig:dens+spin J=1.7}$J=1.7$]{
		\begin{centering}
			\includegraphics[scale=0.36]{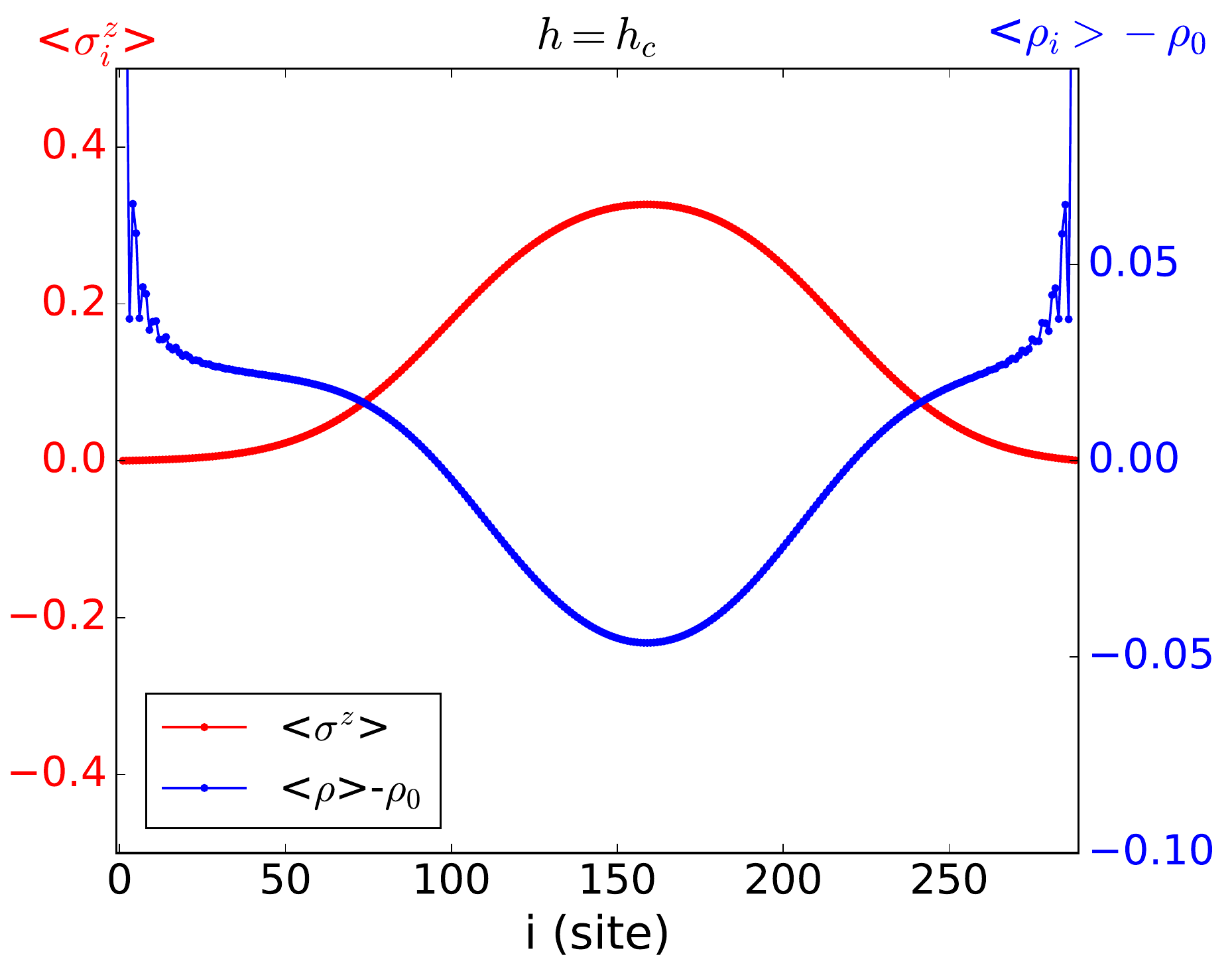} \par
		\end{centering}
		
		}\subfloat[\label{fig:dens+spin J=0.6}$J=0.6$]{
		\begin{centering}
			\includegraphics[scale=0.36]{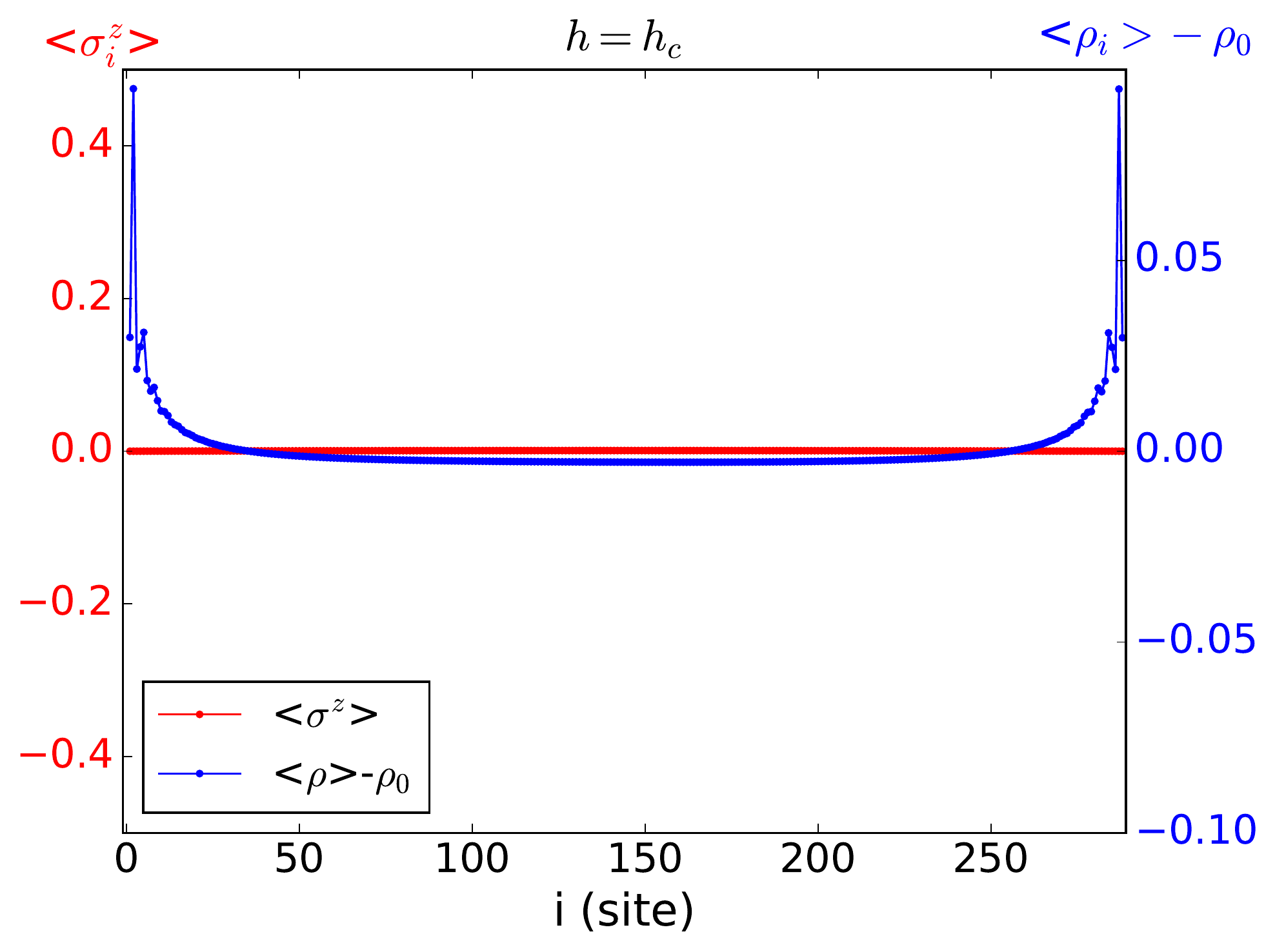} \par
		\end{centering}
		
		} \par
	\end{centering}
	
	\caption{\label{fig:dens+spin profiles}(Color online) Profiles of the density and spin expectation values for different values of $h$ and $J$, showing the onset of phase separation near criticality for $J=1.7$. This was calculated using DMRG for a system with 288 sites, $\tilde{\lambda}=6,t=1, V=0.01$. (a) For $h$ far away from the critical point the profiles are homogenous with no indication of phase seperation. (b) Indication for phase seperation at the critical point, for $J=1.7$. (c) A plot for $J=0.6$ at the critical $h$ showing no phase separation, the shift from average density is only due to boundary effects.} 
\end{figure*}

Having shown indications that the RG flow to strong coupling implies phase separation, we turn to verify this result and to determine how it affects the phase diagram using a numerical calculation in a microscopic model. 
Our starting point for this analysis is a model consisting of an Ising spin chain coupled to an $XXZ$ spin chain described by the Hamiltonian $H=H_{XXZ}+H_{Ising}+H_{Int}$, with
\begin{eqnarray}
	H_{XXZ} & = & -t\sum_{i=1}^{N}\left(S_{i}^{+}S_{i+1}^{-}+h.c\right)-V\sum_{i=1}^{N}S_{i}^{z}S_{i+1}^{z}\label{eq:Hxxz}\\
	H_{Ising} & = & -J\sum_{i=1}^{N}\sigma_{i}^{z}\sigma_{i+1}^{z}-h\sum_{i=1}^{N}\sigma_{i}^{x}\label{eq:HIsing_Lattice}\\
	H_{Int} & = & -{\tilde\lambda}\sum_{i=1}^{N}\left(S_{i}^{z}+\frac{1}{2}\right)\sigma_{i}^{x}\label{eq:H_int_Lattice}.
\end{eqnarray}

In Eq. (\ref{eq:Hxxz}) $S_{i}^{\alpha}$ are spin-$1/2$ operators acting on site $i$. Using a Jordan-Wigner transformation $H_{XXZ}$ can be mapped to a Hamiltonian describing spineless fermions on a lattice with nearest neighbors interaction, where the fermion density at site $i$ is given by $\rho_{i}=S_{i}^{z}+\frac{1}{2}$. At incommensurate filling the long wave length effective theory of $H_{XXZ}$ is given by $H_{LL}$ in Eq. (\ref{eq:H_LL}). 

The Hamiltonian in Eq. (\ref{eq:HIsing_Lattice}) is the transverse field Ising model. Near the critical point at $h=J$, the long wave-length limit of Eq. (\ref{eq:HIsing_Lattice}) is given by the field theory (\ref{eq:Ising H}). 
Finally, the coupling term (\ref{eq:H_int_Lattice}) preserves both the $U(1)$ symmetry of the $XXZ$ model and the $\mathbb{Z}_2$ symmetry of the Ising model. Hence it maps to the  interaction term (\ref{eq:Coupling H}) in the long wavelength limit.

We investigate the model using density matrix renormalization group (DMRG) \cite{White1992} calculations which were performed using the Itensor library \cite{Itensor}.
To tune across the Ising transition we vary the transverse field $h$, while keeping other parameters constant. The critical value $h_c$, depends on the value of the Ising coupling,  but it is not strictly equal to it in the presence of the coupling $\tilde{\lambda}$. By considering the transition tuned by $h$ at different values of $J$ we change the Ising velocity at the critical point and through it the important velocity ratio $u/v$ at the critical point.  In all calculations we use $t=1$, $V=0.01$ and $\rho_{0}=5/18$ with open boundary conditions. We found that it is important to stay sufficiently far from commensurate values of the density, in order to avoid charge density wave instabilities. Also note that the value of V affects the value of the bare Luttinger liquid compressibility in the effective long wave length theory, such that a larger value of V corresponds to a larger value of the bare compressibility. Therefore, choosing relatively small value of V should result in a clearer indication of phase separation at numerically accessible system sizes.

To determine the location of the critical point $h_c(J,\tilde{\lambda})$ we calculate the Binder cumulant $U_{4}\left(h\right)=1-\ex{m^{4}}/\left(3\ex{m^{2}}^{2}\right)$, where $m\equiv\frac{1}{N}\sum_{i}\sigma_{i}^{z}$. The value of $h_{c}$ is obtained from the crossing points of the curves $U_{4}\left(h\right)$ calculated for different system sizes \cite{Binder1981} (see Appendix \ref{sec:locating hc} for details). The value of the Binder cumulant at the critical point $U_{4}^*$ is expected to be universal if the transition is continous. However, in our numerical computation we find that $U_{4}^*$  depends on the ratio J/t, even in the regime where the transition remains second order. This can be attributed to the slow marginal irrelevant flow of the coupling term. 

Having found the transition point, we look for signatures of phase separation, near it as well as deep in the two phases, by examining the profiles of the local density $\ex{\rho_{i}}=\ex{S_{i}^{z}+\frac{1}{2}}$ and of the Ising spin $\ex{\sigma_{i}^{z}}$ along the chain. Examples of calculated profiles are shown in Fig.~\ref{fig:dens+spin profiles}. Away from the critical point (Fig.~\ref{fig:dens+spin J=1.7 not critical}) the profiles are completely flat, showing no indication of phase separation. At the transition we see a big difference between the case with large Ising coupling $J/t=1.7$ (and hence large Ising velocity $u$) shown in Fig.~\ref{fig:dens+spin J=1.7}  compared to the case with smaller Ising coupling $J/t=0.6$ (small Ising velocity) shown in Fig.~\ref{fig:dens+spin J=0.6}. In the former case the density and spin profiles are highly inhomogeneous. The density in the middle of the chain is well below the average while it is above the average in the wings. Correspondingly, the Ising order parameter shows a finite value in the center and is nearly vanishing near the edges. These are clear signatures of phase separation.  In the latter case Fig.~\ref{fig:dens+spin J=0.6}, there is no sign of phase separation even near the critical point. This agrees with the expectation that the Ising critical point survives for sufficiently small ratio $u/v$.

The density and spin profile obtained in the phase separated state can be understood from the effect of the boundaries. Each boundary spin interacts only with one rather than two spins, unlike the bulk spins that have two neighbors. Once phase separation occurs, the weaker effective Ising coupling at the boundaries favors the Ising disordered phase (and correspondingly \emph{higher} density, due to the sign of the coupling term, $\tilde{\lambda}>0$ in Eq. (\ref{eq:H_int_Lattice})). Therefore 
the boundaries act as nucleation centers for the Ising quantum disordered phase when the system becomes susceptible to phase separation close to $h_{c}$. In contrast, as we show below, the small difference between the density near the edge and in the middle in Fig 3(c) ($J/t=0.6$) can be explained as a finite size effect at the critical point. 

\begin{figure}
	\begin{centering}
		\includegraphics[scale=0.45]{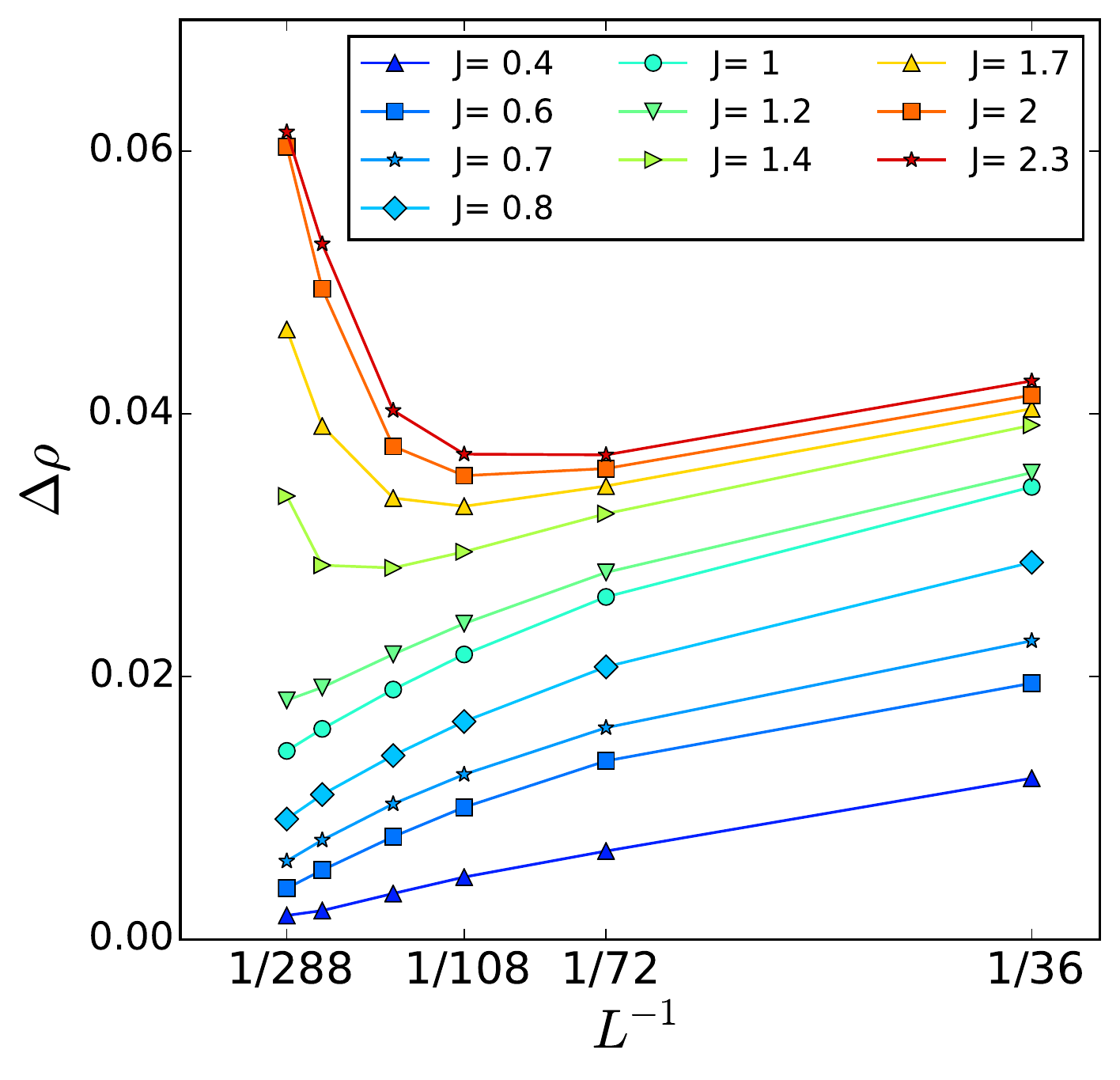} \par
	\end{centering}
	
	\protect\caption{\label{fig:Max phsep}(Color online) Maximal deviation from average density $\Delta\rho$, as defined in Eq. (\ref{eq:deltarho}), calculated for different values of $J$ as a function of inverse system size calculated using $t=1,V=0.01$ and $\tilde{\lambda}=6$. Tuning $J$ in the microscopic model corresponds to tuning the velocity ratio $u/v$ in the resulting effective field theory.  When $J$ is increased there is a transition between a regime in which the maximal density shift decreases with $L$ to a regime in which it increases with $L$. } 
\end{figure}

To determine quantitively whether the system exhibits phase separation for a given value of $J$, we locate the maximal deviation of $\ex{\rho_{i}}$ from the average density $\rho_{0}$ for each value of $h$.  We denote the maximal value of the density shift occurring in the middle third of the system
\begin{equation}
	\Delta\rho\equiv\max_{N/3<i<2N/3}\left|\ex{\rho_{i}}-\rho_{0}\right|\label{eq:deltarho}.
\end{equation}
Note that the position of the minimum can be slightly shifted from the middle of the chain due to slow convergence of DMRG to the ground state in the phase separated regime.

Phase separation is characterized by a finite value of $\Delta\rho$ in the limit $L\to\infty$. On the other hand, if the shift is only due to boundary effects, we expect $\Delta\rho$ to vanish in the thermodynamic limit for all values of $h$. In the latter case we expect $\Delta\rho$ to scale as $1/L$, therefore by performing a linear fit to $\Delta\rho\left(L^{-1}\right)$ we can estimate weather the system is in the phase separated state or not

In Fig. \ref{fig:Max phsep} we plot $\Delta\rho$ as a function of $L^{-1}$ at $h=h_{c}$, for different values of $J$. As $J$ is increased there is a clear transition from a regime in which the system does not exhibit phase separation, where $\Delta\rho$ vanishes linearly with $L^{-1}$, to a regime in which phase separation occurs.
\begin{figure}
	\begin{centering}
		\includegraphics[scale=0.45]{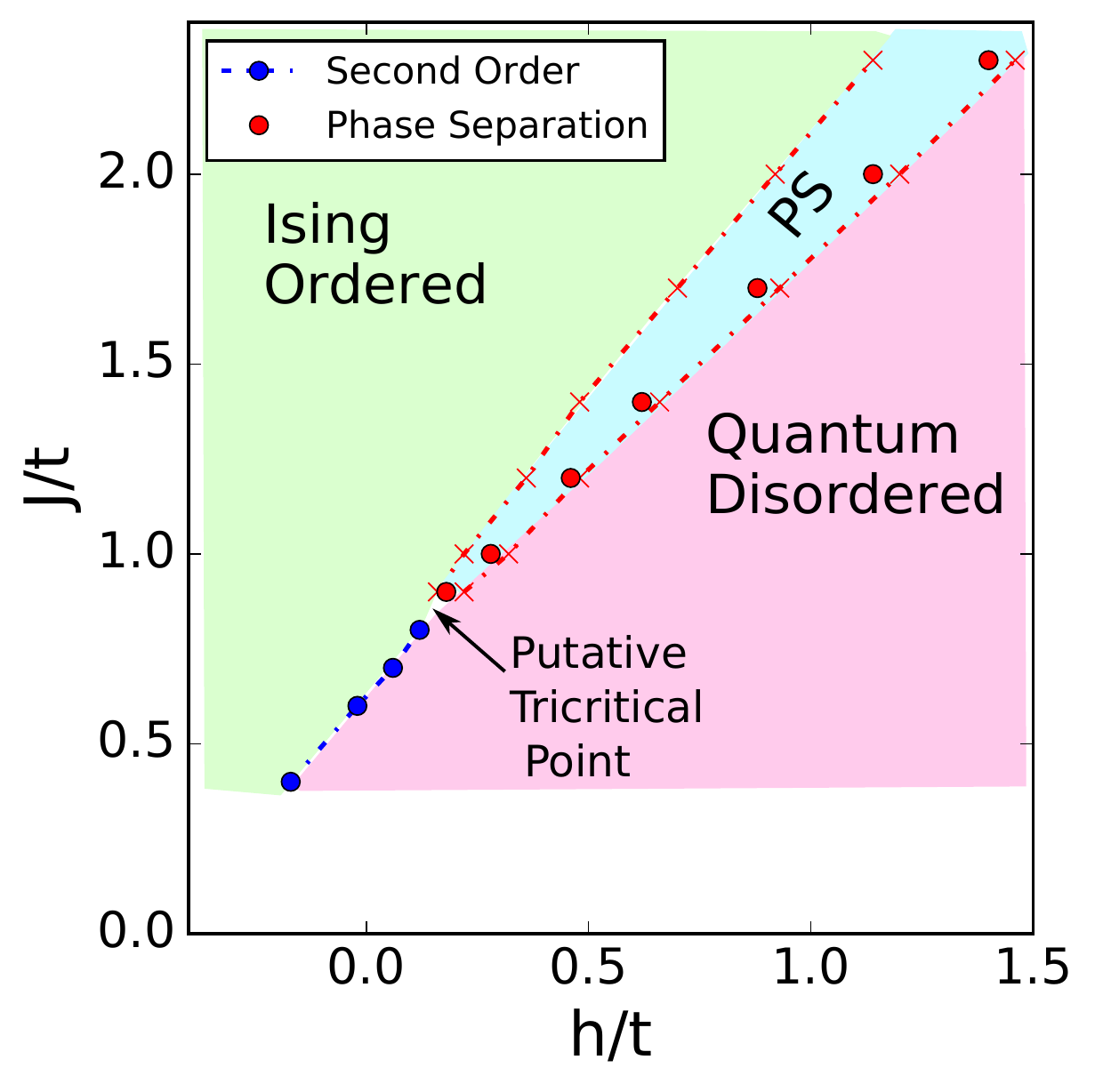} \par
	\end{centering}
	
	\protect\caption{\label{fig:phase diagram}(Color online) Phase diagram of the model described by Eqs. (\ref{eq:Hxxz})-(\ref{eq:H_int_Lattice}), as obtained from a DMRG calculation with $\tilde{\lambda}=6$. The filled circles were obtained from a linear fit to the curves in Fig. (\ref{fig:Max phsep}). The blue circles correspond to curves for which the linear fit is good, the red curves correspond to curves which do not extrapolate to zero linearly. The red crosses are boundaries of the phase separated region, which were estimated from the extrapolation of similar $\Delta\rho\left(L^{-1}\right)$ curves at the corresponding values of h. That is, the left (right) set of crosses is obtained from finding the minimal (maximal) value of h for which the $\Delta\rho\left(L^{-1}\right)$ curves do not extrapolate linearly to zero.} 
\end{figure}

In Fig. \ref{fig:phase diagram} we plot the phase diagram in the $J,h$ plane for a fixed $\tilde\lambda=6$. The values of $J,h$ for which phase separation occurs were obtained by extrapolating the curves $\Delta\rho\left(L^{-1}\right)$ linearly to $L \rightarrow \infty$. Note that although the microscopic value of $\tilde{\lambda}$ might seem large, we still see a regime where the transition is an Ising like second order transition, indicating that the effective coupling is quite weak.

\section{Conclusions\label{sec:summary}}
We investigated the fate of the Ising quantum critical point in one spatial dimension when the Ising degrees of freedom are coupled to a critical bosonic mode. This is a rather generic situation in systems with charge conservation with an internal Ising degree of freedom. Previous one-loop RG calculations \cite{Sitte2009} have identified a regime on the critical manifold in which the Ising critical point is destabilized and the flow is toward strong coupling. Here we find that the slow (logarithmic) flow to strong coupling indicates an instability of the system toward phase separation that manifests only at an emergent length scale, exponentially long in the inverse coupling between the critical modes. This is indicated within the RG calculation by a compressibility that diverges at a finite scale. We also demonstrate the phase separation by direct numerical calculation using DMRG. 

Our result implies the existence of a tricritical point where the line of second order phase transitions terminates and gives way to phase separation (or first order transition in the Grand-canonical picture). An interesting direction for future work is to characterize this tri-critical point, which may exhibit a new class of scaling behavior.
More generally, the model we investigated is the simplest example of non trivial interplay between different critical modes. 
Quantum phase transitions  involving itinerant electrons, where the critical fluctuations of the order parameter are coupled to Fermi-liquid quasiparticles, are important unsolved problems in this class. 
Our one dimensional example demonstrates the role of the \emph{velocity ratio} of the two types of critical modes. Here, as the velocity ratio is varied, the quantum phase transition changes its character from being  continuous to becoming fluctuation-driven first order.

\begin{acknowledgments}
	We would like to thank Liza Huijse and Achim Rosch for useful discussions. EA, JR and OA thank the ERC for financial support through the UQUAM synergy grant. JR acknowledges the support of fellowship from the Gordon and Betty Moore Foundation under the EPiQS initiative (grant no. GBMF4303). EB acknowledges financial support from the European Research Council (ERC) under the European Unions Horizon 2020 research and innovation programme (grant agreement No 639172), and under a Marie Curie CIG grant.
\end{acknowledgments}

\appendix

\section{Calculation of $\eta_B(\ell)$ and $z(\ell)$} 
\label{sec:Calculation of eta_B and z}

In this Appendix we explain how the dynamical exponent $z$ and the phonon field rescaling factor $b^{\eta_B/2}$ are determined. Following the RG scheme of Ref. [\onlinecite{Sitte2009}], as explained in Section~\ref{sec:RG analysis}, we obtain the bosonic part of the action after an infinitesimal RG step
\begin{equation}
\begin{split}
				S_{\text{LL}} = &\int_{-\infty}^{\infty}\d\omega\int_{-\Lambda}^{\Lambda}\d k \frac{b^{\eta_B-3z-1}}{2 \pi vK} \\
				&\phi_{k,\omega}\left(\omega^{2}+b^{2z-2}\left(1-\frac{\pi K}{v}\frac{\partial \Sigma_{ph}}{\partial (k^2)}\bigg|_{k=0}\right)v^{2}k^{2}\right)\phi_{-k,-\omega}
\end{split}
\end{equation}
here $\Sigma_{ph}$ is the one loop phonon self energy calculated from the diagram

\begin{fmffile}{self_energy}
\begin{equation}
\begin{fmfgraph*}(80,50)
	\fmfleft{i}
	\fmfright{o}
	\fmf{dashes}{i,v1}
	\fmf{dashes}{v2,o}
	\fmf{dashes,straight=.8, tension=.3}{v1,v2}
	\fmf{plain,left=.8, tension=.3}{v1,v2}
\end{fmfgraph*}
\end{equation}
\end{fmffile}
where the solid line represents the fermion propagator, the dashed lines represent phonon propagators and the loop integral is evaluated setting the external momenta and frequencies to zero. Thus the one loop self energy is given by 
\begin{equation}
\begin{split}
	\label{eq:self energy}
	\Sigma_{ph}\left(k,\omega\right)&= k^{2}\frac{\lambda^{2}}{2\pi^{4}}\int_{-\infty}^{\infty}\d\nu\int_{\Lambda/b}^{\Lambda}\d p\frac{u^{2}p^{2}+\nu^{2}}{\left(\nu^{2}+\left(up\right)^{2}\right)^{2}} \\
	&=k^{2}\frac{\lambda^{2}}{2\pi^{3}u}\log b,
\end{split}
\end{equation}
to leading order in $k,\omega$.

Requiring that the phonon action remain invariant after the RG step sets the values of $z$ and $\eta_B$
\begin{align}
	&b^{2z-2}\left(1-\frac{\pi K}{v}\frac{\partial \Sigma_{ph}}{\partial (k^2)}\bigg|_{k=0}\right)= 1 \label{eq:eq for z} \\
	&b^{\eta_B-3z-1} =1,
\end{align}
Plugging Eq.~\eqref{eq:self energy} in Eq.~\eqref{eq:eq for z} and exapnding $b=1+\d \ell$ for infinitesimal $\d \ell$, we obtain the values for $z$ and $\eta_B$ given in Eqs.~\eqref{eq: z}-\eqref{eq: eta_B}.
\section{Calculation of $u\left(\ell\right)$\label{sec:Calculation-of l*}}

In this appendix we present the details of the calculation of the flow of $u$ in the $u_{0}/v>1$ case. The first step is to use the fact that $\II\equiv\lambda^{2}Kuv/(u-v)^{4}$ is an invariant of the RG flow which allows us to write Eq. (\ref{eq:du/dl-1}) (without the phenomenological $\lambda^{4}$ term) as 
\begin{equation}
	\frac{
	\partial u}{
	\partial\ell}=-\frac{K\lambda_{0}^{2}u_{0}}{\pi^{2}}\frac{\left(u-v\right)^{4}}{\left(u_{0}-v\right)^{4}}\left[\frac{1}{\left(u+v\right)^{2}}-\frac{1}{4uv}\right]\label{eq:du/dl using I} 
\end{equation}
Integrating (\ref{eq:du/dl using I}) we obtain an implicit equation which determines $u\left(\ell\right)$ 
\begin{equation}
	\ell=\frac{2\pi^{2}v\left(u_{0}-v\right)^{4}}{15K\lambda_{0}^{2}u_{0}}\left[f(u_{0})-f(u)\right]\label{eq:l of u} 
\end{equation}
where $f\left(u\right)=\left(15u^{3}+5u^{2}v+5uv^{2}-v^{3}\right)/(u-v)^{5}$. 

When $u>v$ the function $f\left(u\right)$ is positive and decreasing monotonically with $u$, reaching the value zero in the limit $u\to\infty$. When $\ell$ reaches the value 
\begin{equation}
	\ell^{*}=\frac{2\pi^{2}v\left(u_{0}-v\right)^{4}}{15K\lambda_{0}^{2}u_{0}}f\left(u_{0}\right) 
\end{equation}
Eq. (\ref{eq:l of u}) implies $f\left(u\right)=0$ which happens only when $u$ diverges. 

We can obtain an approximate expression for $u\left(\ell\right)$ in several limiting cases. First, when $u\gg v$, we can approximate $f\left(u\right)\approx15/u^{2}$. Therefore if $u_{0}\gg v$ , since $u$ is increasing with the flow we can plug this approximation in Eq. (\ref{eq:l of u}) and get 
\begin{equation}
	u\left(\ell\right)\approx u_{0}\left(1-\frac{K\lambda_{0}^{2}u_{0}^{3}}{2\pi^{2}v\left(u_{0}-v\right)^{4}}\ell\right)^{-1/2} 
\end{equation}
On the other hand close to the fixed point when $w\equiv1-u/v\ll1$, we can approximate $f\left(u\right)\approx-24v^{-2}w^{-5}$. Therefore in the regime where $w_{0}=1-u_{0}/v\ll1$ we have 
\begin{equation}
	w\approx w_{0}\left(1+\frac{5\II}{16\pi^{2}}w_{0}^{5}\ell\right)^{-1/5}\label{eq:w(l)} 
\end{equation}
expressing the exponent in Eq. (\ref{eq:comp as function of u}) in terms of $w$ and plugging in Eq. (\ref{eq:w(l)}) in the limit of large $\ell$ we obtain Eq. (\ref{eq:chi close to fixed point}).

\section{Finite Size Scaling of $U_{4}$ \label{sec:locating hc}}

\begin{figure}
	\begin{centering}
		\includegraphics[scale=0.36]{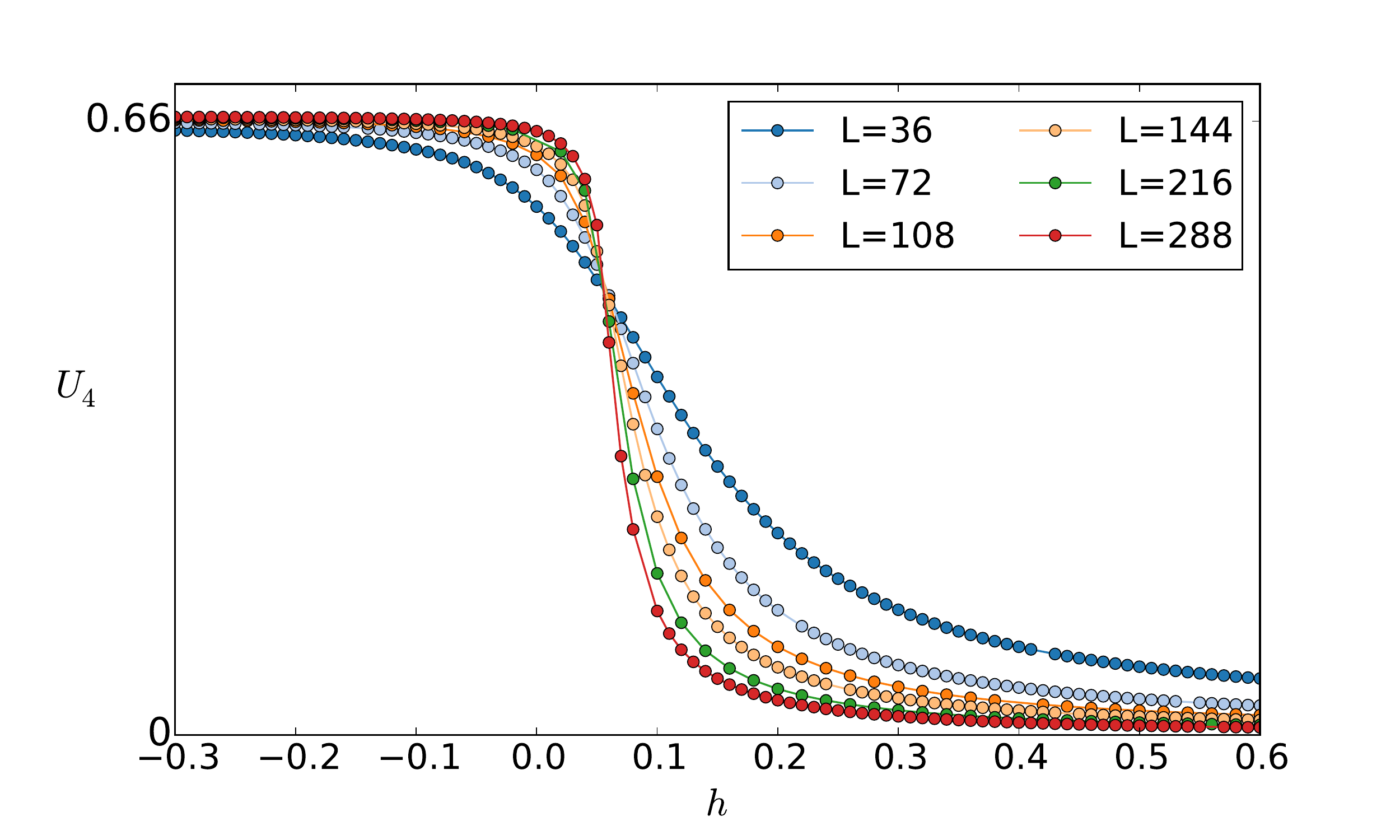} \par
	\end{centering}
	
	\protect\caption{\label{fig:binder}(Color online) Binder cumulant $U_{4}$ as a function of transverse field strength h, calculated for J/t=0.7.
	The different curves are for different system sizes. The lines are interpolation between the data points.} 
\end{figure}

In this appendix we present finite size scaling data of the Binder cumulant which was used to determine the critical value of the transverse field $h$ in the microscopic model described by Eqs. \eqref{eq:Hxxz} -\eqref{eq:H_int_Lattice}. 
Plot of the Binder cumulant computed for different system sizes at $J/t=0.7$ is shown in Fig. \ref{fig:binder}. 
If the transition is second order one expects the Binder comulant to obey the finite size scaling form 
\begin{equation}
	\label{eq:U4 FFS}
		U_{4}\left( h,L \right)=\mathcal{F}\left(\left| h-h_{c} \right|L^{1/\nu}\right)
\end{equation}
where $\mathcal{F}$ is a universal scaling function. In particular Eq. \eqref{eq:U4 FFS} implies that $U_{4}\left( h_{c},L \right)$ is independent of system size. Therefore the value of $h_{c}$ can be obtained from the crossing points of the curves corresponding to different system sizes. Since there are subleading finite size corrections to the scaling form in Eq. ~\eqref{eq:U4 FFS} we obtain a series of crossing points which converges to the value of $h_{c}$ with increasing system size.

\begin{figure}
	\begin{centering}
		\includegraphics[scale=0.36]{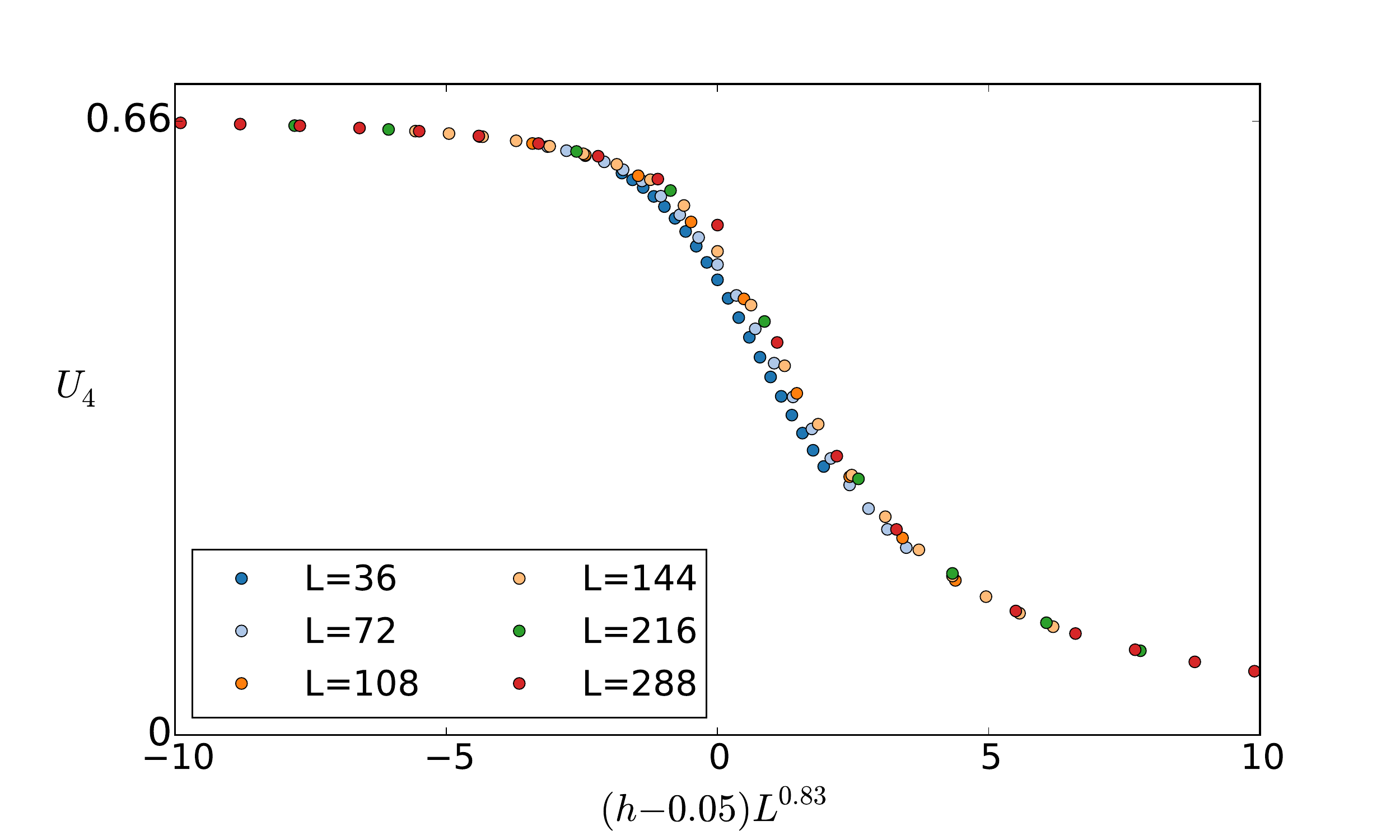} \par
	\end{centering}
	
	\protect\caption{\label{fig:Binder_0.7_nu}(Color online) Binder cumulant $U_{4}$ plotted as a function the rescaled $h$ axis, calculated for J/t=0.7. There is no value of $\nu$ for which the data collapses to a single curve} 
\end{figure}

In Fig. ~\ref{fig:Binder_0.7_nu} we try to obtain data collapse according to the scaling form \eqref{eq:U4 FFS}. This is done for the data points in the interval $\left| h-h_{c} \right|<0.1$. We are not able to obtain good data collapse by fitting the value of $\nu$. To understand this one has to remember that  the scaling form in Eq. \ref{eq:U4 FFS} relies on the assumption that close to the critical point the correlation length $\xi$ diverges as $\xi \sim \left| h-h_{c} \right|^{-\nu}$. However, when a marginal operator is present (the coupling term $\lambda$ in our case) it can lead to corrections of the correlation length scaling.

\begin{figure}
	\begin{centering}
		\includegraphics[scale=0.36]{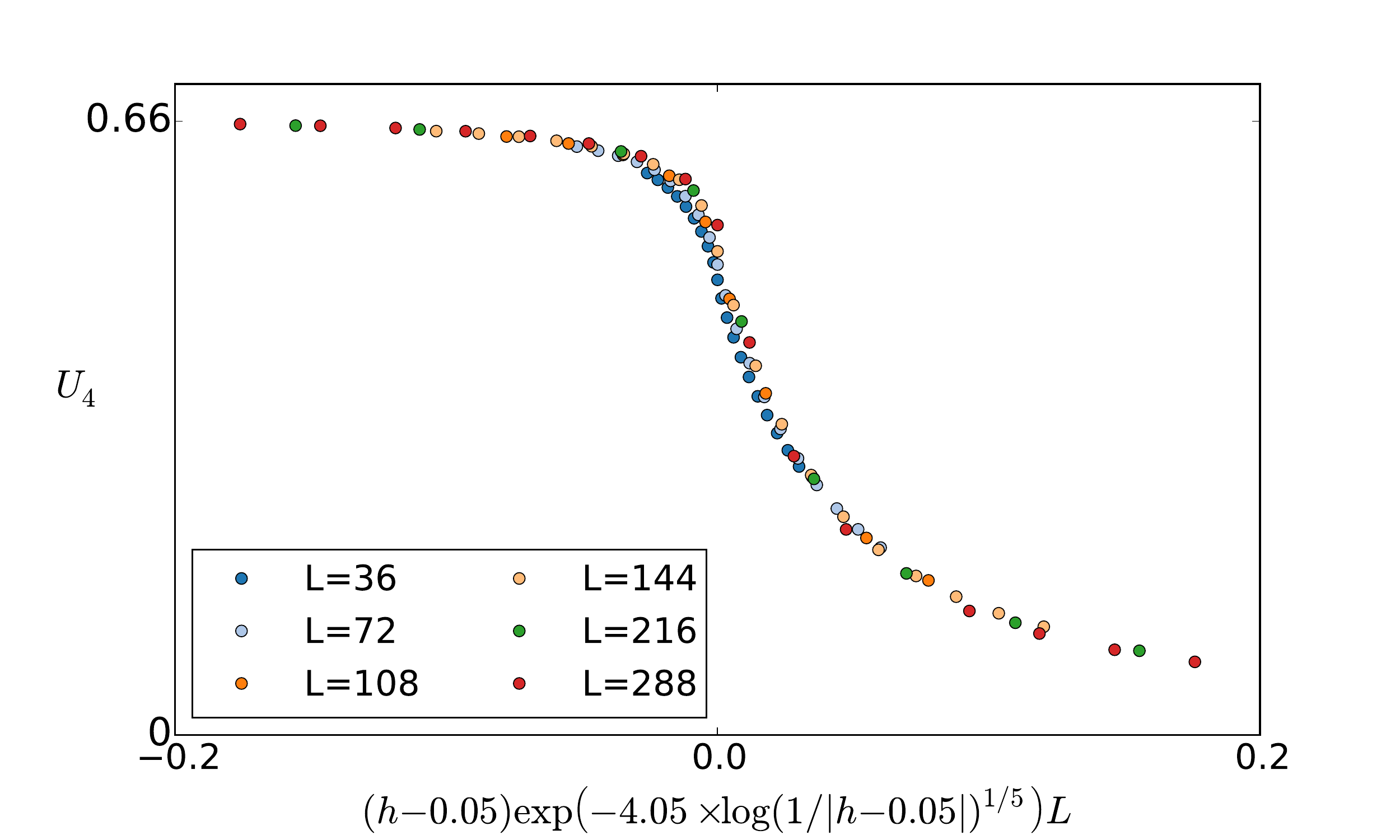} \par
	\end{centering}
	
	\protect\caption{\label{fig:Binder_0.7_exp}(Color online) Binder cumulant $U_{4}$ plotted as a function of $L/\xi(h)$ (where $\xi(h)$ is defined in Eq. \eqref{eq: xi corrected} ), calculated for J/t=0.7. We are able to fit a value of $\alpha$ for which there is data collapse. } 
\end{figure}

It was shown in Ref. [\onlinecite{Sitte2009}] that in the regime where the transition is second order the Ising gap is suppressed due to the slow flow of the coupling term. This implies that the divergence of correlation length is enhanced, since the correlation length is related to the Ising gap by $\xi=u/\Delta$. In particular we find that close to the critical point 
\begin{equation}
	\label{eq: xi corrected}
	\xi\sim \frac{1}{\left| h-h_{c} \right|}\exp\left(2\left(\alpha\log\left(\frac{1}{\left| h-h_{c} \right|}\right)\right)^{1/5}\right)
\end{equation}
where $\alpha \equiv 5\II/(16\pi^{2})$ is a non universal factor which depends on the bare parameters $v,K,\lambda,u$. In Fig. \ref{fig:Binder_0.7_exp} we show data collapse when the Binder cumulant data is plotted as a function of $L/\xi(h)$, with $\xi(h)$ defined from Eq. \eqref{eq: xi corrected}. We can see better data collapse assuming the corrected scaling of $\xi$.

\FloatBarrier

\bibliographystyle{apsrev4-1} 
%

\end{document}